\pdfoutput=1
\documentclass{article}
\usepackage{spconf,amsmath,graphicx,siunitx,etoolbox,booktabs,hyperref,multirow}
\usepackage[capitalize]{cleveref}
\usepackage[table]{xcolor}
\usepackage[acronym]{glossaries}
\glsdisablehyper
\usepackage{balance}
\usepackage{tikz}
\usetikzlibrary{positioning,arrows,calc}
\usepackage{pifont}
\usepackage{tabularx}

\tikzset{
    box/.style={draw,black,thick},
    arrow/.style={draw,black,->,>=stealth,shorten >=.3mm},
    branch/.style={inner sep=0.3mm,circle,fill=black}
}

\newcommand{\mrow}[1]{\multirow{2}[2]{*}{\begin{tabular}{@{}c@{}}#1\end{tabular}}}
\newcommand{\cmark}{\ding{51}}

\newcommand{\inred}[1]{\textcolor{red}{#1}}
\def\wsjzerotwomix{WSJ0-2mix}
\def\wsjtwomix{WSJ-2mix}

\newcommand{\vect}{\ensuremath{\mathbf}}
\def\L{\ensuremath{\mathcal{L}}}
\def\Lfe{\ensuremath{\L^{(\mathrm{FE})}}}
\def\Lasr{\ensuremath{\L^{(\mathrm{ASR})}}}

\newcolumntype{H}{>{\setbox0=\hbox\bgroup}c<{\egroup}@{}}   
\newcommand{\tbli}{\phantom{+}}
\newcommand{\tblii}{\phantom{++}}

\robustify\inred

\newacronym{WER}{WER}{Word Error Rate}
\newacronym{CER}{CER}{Character Error Rate}
\newacronym{SDR}{SDR}{Signal-to-Distortion Ratio}
\newacronym{E2E}{E2E}{End-to-End}
\newacronym{SOTA}{SOTA}{State of the Art}
\newacronym{ASR}{ASR}{Automatic Speech Recognition}
\newacronym{SI-SNR}{SI-SNR}{Scale-Invariant-Signal-to-Distortion Ratio}
\newacronym{DPCL}{DPCL}{Deep Clustering}
\newacronym{PIT}{PIT}{Permutation Invariant Training}
\newacronym{STFT}{STFT}{Short-Time Fourier Transform}

\title{End-to-end training of time domain audio separation and recognition}
\name{\begin{tabular}{c}
     Thilo von Neumann$^{1,2}$ \qquad Keisuke Kinoshita$^{1}$ \qquad Lukas Drude$^{2}$ \qquad Christoph Boeddeker$^{2}$ \\
     Marc Delcroix$^{1}$ \qquad Tomohiro Nakatani$^{1}$ \qquad Reinhold Haeb-Umbach$^{2}$
\end{tabular}}
\address{$^{1}$NTT Communication Science Laboratories, NTT Corporation, Kyoto, Japan \\ $^{2}$Paderborn University, Department of Communications Engineering, Paderborn, Germany}

\begin{document}
\ninept
\maketitle
\begin{abstract}
The rising interest in single-channel multi-speaker speech separation sparked development of \gls{E2E} approaches to multi-speaker speech recognition.
However, up until now, state-of-the-art neural network--based time domain source separation has not yet been combined with \gls{E2E} speech recognition.
We here demonstrate how to combine a separation module based on a Convolutional Time domain Audio Separation Network (Conv-TasNet) with an \gls{E2E} speech recognizer and how to train such a model jointly by distributing it over multiple GPUs or by approximating truncated back-propagation for the convolutional front-end.
To put this work into perspective and illustrate the complexity of the design space, we provide a compact overview of single-channel multi-speaker recognition systems.
Our experiments show a word error rate of \SI{11.0}{\percent} on \wsjzerotwomix{} and indicate that our joint time domain model can yield substantial improvements over cascade DNN-HMM and monolithic \gls{E2E} frequency domain systems proposed so far.
\end{abstract}
\vspace{-1.5mm}
\section{Introduction}
\label{sec:intro}
\vspace{-1.5mm}
%
%
%
\gls{ASR} is a key technology for the task of automatic analysis of any kind of spoken speech, e.g., phone calls or meetings.
For scenarios of relatively clean speech, e.g., recordings of telephone speech or audio books, \gls{ASR} technologies have improved drastically over the recent years \cite{Xiong2018_Microsoft2017ConversationalSpeech}.
More realistic scenarios like spontaneous speech or meetings with multiple participants in many cases require the \gls{ASR} system to recognize the speech of multiple speakers simultaneously.
In meeting scenarios for example, the overlap is in the range of \SIrange{5}{10}{\percent}\footnote{Measured on the AMI meeting corpus \cite{ami}.} and can easily exceed \SI{20}{\percent} in informal get-togethers\footnote{Measured on the CHiME-5 database.}.
Thus, there has been a growing interest in source separation systems and multi-speaker \gls{ASR}.
A special focus lies on processing of single-channel recordings, as this is not only important in scenarios where only a single-channel is available (e.g., telephone conference recordings), but as well for multi-channel recordings where conventional multi-channel processing methods, e.g., beamforming, cannot separate the speakers well enough in case, e.g., they are spatially too close to each other.

%
%

The topic of single-channel source separation has been examined extensively over the last few years, trying to solve the cocktail party problem with techniques such as \gls{DPCL} \cite{Isik2016_Singlechannelmulti}, \gls{PIT} \cite{Yu2016_PermutationInvariantTraining} and TasNet \cite{Luo2017_tasnet,Luo2018_ConvTasNetSurpassing}.
In \gls{DPCL}, a neural network is trained to map each time-frequency bin to an embedding vector in a way that embedding vectors of the same speaker form a cluster in the embedding space.
These clusters can be found by a clustering algorithm and be used for constructing a mask for separation in frequency domain.
Concurrently, \gls{PIT} has been developed which trains a simple neural network with multiple outputs to estimate a mask for each speaker with a permutation invariant training criterion.
The reconstruction loss is calculated for each possible assignment of training targets to estimations for a mixture, and the permutation that minimizes the loss is then used for training.
Both \gls{DPCL} and \gls{PIT} show a good separation performance in time-frequency domain.
The permutation-invariant training scheme was adopted to time domain source separation with a Time domain Audio Separation Network (TasNet) which replaces the commonly used \gls{STFT} with a learnable transformation and directly works on the raw waveform.
TasNet achieves a \gls{SDR} gain of more than \SI{15}{\decibel} on \wsjzerotwomix, even outperforming oracle masking in frequency domain.

%
%

Based on these source separation techniques, multi-speaker \gls{ASR} systems have been constructed.
\gls{DPCL} and \gls{PIT} have been used as frequency domain source separation front-ends for a state-of-the-art single-speaker \gls{ASR} system and extended to jointly trained \gls{E2E} or hybrid systems \cite{Menne2019_AnalysisDeepClustering,Settle2018_Endendmulti,Yu2017_RecognizingMultitalker,Qian2018_Singlechannelmulti}.
They showed that joint (re-)training can improve the performance of these models over a simple cascade system.
The effectiveness of TasNet as a time domain front-end for \gls{ASR} was investigated in \cite{Bahmaninezhad2019_comprehensivestudyspeech}, showing an improvement over frequency domain processing for both source separation and \gls{ASR} results.
However, TasNet was not yet optimized jointly with an \gls{ASR} system, possibly due to the intricacies of dealing with the high memory consumption or the novelty of the TasNet method.

%
%

In this paper, we combine a state-of-the-art front-end, i.e., Conv-TasNet [4], with an \gls{E2E} CTC/attention \cite{Kim2017_JointCTCattention,Watanabe2017_HybridCTC/attentionarchitecture,Chan2016_Listenattendspell} \gls{ASR} system to form
an  E2E  multi-speaker  \gls{ASR}  system  that  directly  operates  on  raw
waveform  features.
We try to answer the questions whether it is possible to jointly train a time domain source separation system like Conv-TasNet with an \gls{E2E} \gls{ASR} system and whether the performance can be improved by joint fine-tuning.
Going further on the investigations from \cite{Bahmaninezhad2019_comprehensivestudyspeech}, we retrain pre-trained front- and back-end models jointly and show by evaluating on the \wsjzerotwomix{} database that a simple combination of an independently trained Conv-TasNet and \gls{ASR} system already provides competitive performance compared to other \gls{E2E} approaches, while a joint fine-tuning of both modules in the style of an \gls{E2E} system can further improve the performance by a large margin.
We enable joint training by distributing the model over multiple GPUs and show that an approximation of truncated back-propagation~\cite{werbos1990backpropagation} through time for convolutional networks enables joint training even on a single GPU by significantly reducing the memory usage while still providing a good performance.

%
%

We finally put this work into perspective by providing a compact overview of single-channel multi-speaker \gls{ASR} system and illustrate the complexity of the design space.

\begin{figure*}[t]
    \centering
    \begin{tikzpicture}
    \node[box,minimum height=2cm,minimum width=10mm,fill=black!10] (convtas) {};
    \node[box,minimum height=2cm,minimum width=10mm,right=20mm of convtas] (perm) {};
    
    \coordinate (perm-in-1) at ($(perm.west)+(0,7mm)$);
    \coordinate (perm-in-2) at ($(perm.west)+(0,-7mm)$);
    \coordinate (perm-out-1) at (perm.east|-perm-in-1);
    \coordinate (perm-out-2) at (perm.east|-perm-in-2);
    \node[box,right=3cm of perm-out-1,minimum height=1cm,fill=black!10] (asr1) {\begin{tabular}{c}
        ASR \\ encoder
    \end{tabular}};
    \node[box,right=3cm of perm-out-2,minimum height=1cm,fill=black!10] (asr2) {\begin{tabular}{c}
        ASR \\ encoder
    \end{tabular}};
    \node[branch,right=5mm of asr1] (br3) {};
    \node[branch,right=5mm of asr2] (br4) {};
    \node[box,above right=.5mm and 10mm of asr1.east,minimum height=.5cm,fill=green!10] (ctc1) {CTC};
    \node[box,above right=.5mm and 10mm of asr2.east,minimum height=.5cm,fill=green!10] (ctc2) {CTC};
    \node[box,below right=.5mm and 10mm of asr1.east,minimum height=.5cm,fill=blue!10] (dec1) {att. decoder};
    \node[box,below right=.5mm and 10mm of asr2.east,minimum height=.5cm,fill=blue!10] (dec2) {att. decoder};

    \node[box,fill=orange!10, yshift=-17.5mm] (sisnr) at ($(convtas.east)!1/2!(perm.west)$) {\begin{tabular}{c} perm. inv. \\ SI-SNR loss \end{tabular}};

    \coordinate (mix) at ($(convtas.west) + (-10mm, 0)$);
    \coordinate (sin1) at ($(convtas.west|-sisnr.west) + (-10mm, 3mm)$);
    \coordinate (sin2) at ($(convtas.west|-sisnr.west) + (-10mm, -3mm)$);

    \node (mix-audio) [anchor=east] at (mix) {\includegraphics[height=1cm]{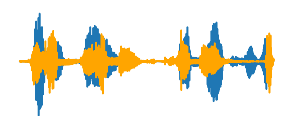}};
    \node (s1-audio)  [anchor=east] at (sin1) {\includegraphics[height=1cm]{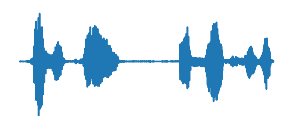}};
    \node (s2-audio)  [anchor=east] at (sin2) {\includegraphics[height=1cm]{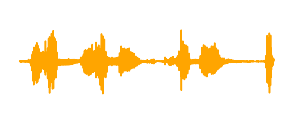}};

    \node[right=1cm of dec1] (l-ctc) {$\L^{\mathrm{(CTC)}}$};
    \node[] at (ctc2-|l-ctc) (l-att) {$\L^{\mathrm{(att)}}$};
    \node[right=2cm of sisnr] (l-fe) {\Lfe};
    
    \draw[arrow] (mix) -- node [above] {$\vect{x}$} (convtas);
    \draw[] (perm-in-1) -- (perm-out-1);
    \draw[] (perm-in-1) -- (perm-out-2);
    \draw[] (perm-in-2) -- (perm-out-2);
    \draw[] (perm-in-2) -- (perm-out-1);
    \draw[] (convtas.east|-perm-in-1) --node[branch,pos=0.33] (br1) {} (perm-in-1);
    \draw[] (convtas.east|-perm-in-2) --node[branch,pos=0.66] (br2) {} (perm-in-2);
    \draw[arrow] (br1) -- (br1|-sisnr.north);
    \draw[arrow] (br2) -- (br2|-sisnr.north);
    \draw[arrow] (sisnr) -|node[right]{$\pi_\mathrm{sig}$} (perm);
    \draw[arrow] (perm-out-1) -- node [above,near start] {$\vect{x}^{\mathrm{(enh)}}_1$} node[above, at end, anchor=south east] {\includegraphics[height=.7cm]{img/s1.png}} (asr1);
    \draw[arrow] (perm-out-2) -- node [above,near start] {$\vect{x}^{\mathrm{(enh)}}_2$} node[above, at end, anchor=south east] {\includegraphics[height=.7cm]{img/s2.png}} (asr2);
    \draw[arrow] (sin1) -- node [above] {$\vect{x}_1$} +(10mm, 0) -- (sisnr.west|-sin1);
    \draw[arrow] (sin2) -- node [above] {$\vect{x}_2$} +(10mm, 0) -- (sisnr.west|-sin2);
    \draw (asr1) -- (br3);
    \draw (asr2) -- (br4);
    \draw[arrow] (br3) |- (ctc1);
    \draw[arrow] (br3) |- (dec1);
    \draw[arrow] (br4) |- (ctc2);
    \draw[arrow] (br4) |- (dec2);
    \draw[] (mix-audio) -- (mix);
    \draw[] (s1-audio) -- (sin1);
    \draw[] (s2-audio) -- (sin2);

    \draw[arrow,very thick,green!40,dashed] (ctc1) to[out=0,in=170] (l-ctc);
    \draw[arrow,very thick,green!40,dashed] (ctc2) to[out=0,in=190] (l-ctc);
    \draw[arrow,very thick,blue!40,dashed] (dec1) to[out=0,in=170] (l-att);
    \draw[arrow,very thick,blue!40,dashed] (dec2) to[out=0,in=190] (l-att);
    \draw[arrow,very thick,dashed,orange!45] (sisnr) to[out=350,in=200] (l-fe);

    \node[fill=white, inner sep=0,rotate=90] at (perm) {perm. assign.};
    \node[inner sep=0,rotate=90] at (convtas) {Conv-TasNet};

    \end{tikzpicture}
    \caption{Architecture of the joint E2E ASR model. Sources are separated by a Conv-TasNet and separated audio streams are processed by a single-speaker ASR system. During training, the permutation problem is solved based on the signal level loss with $\pi_\mathrm{sig}$.}
    \label{fig:architecture}
\end{figure*}
    
\section{Relation to Prior Work}
\label{sec:related}
%
%
Other works already studied the effectiveness of frequency domain source separation techniques as a front-end for \gls{ASR}.
\gls{DPCL} and \gls{PIT} have been efficiently used for this purpose, and it was shown that joint retraining for fine-tuning can improve performance \cite{Menne2019_AnalysisDeepClustering,Settle2018_Endendmulti,Qian2018_Singlechannelmulti}.
%
%
\gls{E2E} systems for single-channel multi-speaker \gls{ASR} have been proposed that  no longer consist of individual parts dedicated for source separation and speech recognition, but combine these functionalities into one large monolithic neural network.
They extend the encoder of a CTC/attention-based \gls{E2E} \gls{ASR} system to separate the encoded speech features and let one or multiple attention decoders generate an output sequence for each speaker \cite{Seki2018_purelyendend,Chang2019_EndEndMonauralMulti}.
These models show promising performance, but they are not on par with hybrid cascade systems yet.
Drawbacks of these monolithic \gls{E2E} models compared to cascade systems include that they cannot make use of parallel and single-speaker data and that they do not allow pre-training of individual system parts.
%
%
The impact of using raw waveform features directly for the task of multi-speaker \gls{ASR} has only been investigated for a combination of TasNet and a single-speaker \gls{ASR} system \cite{Bahmaninezhad2019_comprehensivestudyspeech}, but not yet jointly trained.

\section{Source separation and speech recognition}
\label{sec:proposed}
\subsection{Time domain source separation with Conv-TasNet}
Conv-TasNet \cite{Luo2018_ConvTasNetSurpassing} is a single-channel source separating front-end which can produce waveforms for a fixed number of speakers from a mixture waveform.
It is a variant of \cite{Yu2016_PermutationInvariantTraining}, replacing the feature extraction by a learnable transformation and the separation network by a convolutional architecture.
It outputs an estimated audio stream in time domain for each speaker present in the input signal $\vect{x}$:
\begin{equation}
    [\vect{x}^{\mathrm{(enh)}}_1, \vect{x}^{\mathrm{(enh)}}_2] = \mathrm{ConvTasNet}(\vect{x}).
\end{equation}
The model directly works on the raw waveform instead of STFT frequency domain features which makes it possible both to easily model and reconstruct phase information and to propagate gradients through the feature extraction and signal reconstruction parts.
While propagating from raw waveform at the output to raw waveform at the input, it is possible to directly optimize a loss on the time domain signals, such as the \gls{SI-SNR} loss, which we call the front-end loss \Lfe here.
This loss is optimized in a permutation-invariant manner by picking the assignment $\pi_\mathrm{sig}$ of estimations to targets that minimizes the loss.

Since the Conv-TasNet is built upon a convolutional architecture, it can be heavily parallelized on GPUs as compared to RNN-based models, but has a limited receptive field.
When optimized for source separation only, the limited length of the receptive field is actively exploited by performing a chunk-level training on chunks of \SI{4}{\second} length randomly cut from the training examples, both increasing the variability of data within one minibatch and simplifying implementation, since the length of all training examples is fixed.

\subsection{End-to end CTC/attention speech recognition}

As a speech recognizer we use a CTC/attention-based \gls{ASR} system.
We use an architecture similar to \cite{Watanabe2017_HybridCTC/attentionarchitecture} with an implementation included in the ESPnet framework \cite{Watanabe2018_ESPnet}, but we replace the original filterbank and pitch feature extraction by log-mel features implemented in a way such that gradients can be propagated through.
This way, gradients can flow from the \gls{ASR} system to the front-end.

The multi-target loss for the ASR system \Lasr{} is composed of a CTC and an attention loss, as in \cite{Watanabe2017_HybridCTC/attentionarchitecture} Sections II B and C,
\begin{equation}
    \Lasr = \lambda \L^{(\mathrm{CTC})} + (1 - \lambda)\L^{(\mathrm{att})}
\end{equation}
with a weight $\lambda$ that controls the interaction of both loss terms.
%
%
During training, teacher forcing using the ground truth transcription labels is employed for the attention decoder.

\section{Joint End-to-end multi-speaker asr}
%
%
We propose to combine a Conv-TasNet as the source separation front-end with a CTC/attention speech recognizer as displayed in \cref{fig:architecture}.
The input mixture $\vect{x}$ is separated by the front-end and the separated audio streams are processed by a single-speaker \gls{ASR} back-end.
%
%
Although multi-speaker source separation can already be performed by combining independently trained front- and back-end systems, the source separator produces artifacts unknown to the \gls{ASR} system which disturb its performance.
According to \cite{Heymann2017_BeamnetEndend}, and as also shown in \cite{Settle2018_Endendmulti,Qian2018_Singlechannelmulti}, such a mismatch can be mitigated by jointly fine-tuning the whole model at once.

%
%
We here compare three different variants of joint fine-tuning: (a) fine-tuning just the ASR system on the enhanced signals, (b) fine-tuning just the front-end by propagating gradients through the ASR system but only updating the front-end parameters and (c) jointly fine-tuning both systems.
%
%
The losses for the front- and back-end are combined as
\begin{equation}
    \L = \alpha\Lfe + \beta\Lasr,    
\end{equation}
where $\alpha$ and $\beta$ are manually chosen weights for the front-end and \gls{ASR} losses.
For (a) $\alpha$ is set to $0$ and $\beta$ to $1$, for (b) $\alpha$ is set to $1$ and $\beta$ to $0$, and for (c) $\beta$ is set to $1$ and $\alpha$ is set to $0.5$ without carefully optimizing them.
The system does not seem to be very sensitive to the choices of $\alpha$ and $\beta$.

%
%

In order to choose the transcription for teacher forcing and loss computation a permutation problem needs to be solved.
Two possible options are to use the permutation $\pi_\mathrm{CTC}$ that minimizes the CTC loss as in \cite{Seki2018_purelyendend}, or the permutation $\pi_\mathrm{sig}$ that minimizes the signal level loss, as in \cite{Settle2018_Endendmulti}.
While using $\pi_\mathrm{CTC}$ has the advantage of not requiring parallel data, permutation assignment based on $\pi_\mathrm{sig}$ works more reliably in our experiments and we use $\pi_\mathrm{sig}$ for all fine-tuning experiments even when the front-end is not optimized.

\subsection{Approximated truncated back-propagation through time}
%
%

One-dimensional Convolutional Neural Networks (1D-CNNs) over time, as they are used in the Conv-TasNet, can be seen as an alternative to Recurrent Neural Network (RNN) architectures.
Similar to RNNs, this can lead to enormous memory consumption when a sufficiently long time series is used for back-propagation.
For example, we here fine-tune the Conv-TasNet with the E2E \gls{ASR} model jointly on single mixtures.
Although we constrain ourselves to a batch size of one, this requires us to split the model onto four GPUs, three GPUs for the front-end and one GPU for the back-end, by placing individual layers on different devices.

This memory consumption can be addressed by generalizing truncated back-propagation through time (TBPTT) to 1D-CNN architectures.
TBPTT for 1D-CNNs can in theory be realized by back-propagating the gradients for a part of the output only.
While moving back towards the input, the gradients reaching over the borders of this part are ignored.
In practice, however, this is difficult to implement and we here approximate TBPTT for the Conv-TasNet front-end by ignoring the left and right context of the block the gradients are computed for.
We first compute the forward step on the whole mixture without building the backward graph to obtain an output estimation for the whole signal.
Note that this only requires to store the output signal and no persistent data for the backward computation.
We then compute the forward step again with enabled backward graph construction, but only for a chunk randomly cut from the input signal.
The approximated output for the whole utterance is formed by overwriting the corresponding part of the full forward output with the approximated chunk output.
This full output is passed to the back-end and gradients reaching the front-end are only back-propagated through the approximated chunk.
This technique allows to run the joint training on a single GPU in our case, but even with larger GPU memory this permits increasing the batch size which in general speeds up training and produces a smoother gradient.

\section{Experiments}
\label{sec:experiments}
%

%
%
We carry out experiments on the WSJ database and the commonly used \wsjzerotwomix{} dataset first proposed in \cite{Isik2016_Singlechannelmulti}.
The data in \wsjzerotwomix{} is generated by linearly mixing clean utterances from WJS0 (si\_tr\_s for training and si\_et\_05 for testing) at ratios randomly chosen from \SIrange{0}{5}{\decibel}.
%
%
It consists of two different datasets, namely the min and max datasets.
The min dataset was designed for source separation and is formed by truncating the longer one of the two mixed recordings, so that it only contains fully overlapped speech.
We use this dataset for pre-training the Conv-TasNet, but it is not suitable for joint training where the audio data needs to match the full transcription.
For this need, we use the max dataset that does not truncate any recordings.
%
%
We use a sampling frequency of \SI{8}{kHz} for both the front- and back-end to speed up the training process.
%
%
We remove any labels marked as noisy, i.e., special tokens such as "lip smack" or "door slam", from the training transcriptions since these cannot be assigned to one speaker based on speech information by the front-end which makes their estimation ambiguous.

%
We evaluate our experiments in terms of \gls{WER}, \gls{CER} and, where applicable, by \gls{SDR} as supplied by the BSS-EVAL toolbox \cite{Fevotte2005_BSS_EVAL} and \gls{SI-SNR} \cite{LeRoux2019_SDR}.
For the experiments on mixed speech, the metrics are computed for all possible combinations of predictions and ground truth transcriptions and the metrics for the permutation that minimizes \gls{WER} is reported.


\subsection{Conv-TasNet time domain source separation}
%
%
We use the best performing architecture according to \cite{Luo2018_ConvTasNetSurpassing} and optimize it with the ADAM optimizer \cite{Kingma2014_Adam}.\footnote{Our implementation is based on https://github.com/funcwj/conv-tasnet.}
In particular, following the hyper-parameter notations in the original paper, we set $N=512$, $L=16$, $B=128$, $H=512$, $P=3$, $X=8$ and $R=3$ with global layer normalization.
For distributing over multiple GPUs, we split between the three repeating convolutional blocks.
%
%
\cref{tab:tas} lists SDR and \gls{SI-SNR} performance for our Conv-TasNet model, comparing the min and max subsets of \wsjzerotwomix.
It can be seen that our implementation of the Conv-TasNet reaches a comparable performance on the min dataset when compared to the original paper.
%
%
There is a slight degradation in performance on the max dataset caused by the mismatch of training and test data because the model never saw long single-speaker regions during training and learned to always output a speech signal on both outputs, while these regions are present in the max dataset.

\begin{table}[ht]
\vspace{-1em}
    \centering
    \caption{SDR and SI-SNR in \si{\decibel} for the min and max test (tt) datasets of the \wsjzerotwomix{} database. 
    }
    \label{tab:tas}
    \begin{tabular}{lSS}
        \toprule
         Dataset & {SDR} & {SI-SNR} \\
         \midrule
         \wsjzerotwomix{} min \cite{Luo2018_ConvTasNetSurpassing} & 15.6 & 15.3 \\
         \wsjzerotwomix{} min (ours) & 14.3 & 13.8 \\
         \wsjzerotwomix{} max (ours) & 13.8 & 13.4 \\
         \bottomrule
    \end{tabular}
    \vspace{-1.5em}
\end{table}
\begin{table*}[t]
    \def\ditto{''}  
    \def\thl{\fontseries{b}\selectfont}
    \robustify\fontseries
    \robustify\selectfont
    \centering
    \caption{CER and WER on max test (tt) set of \wsjzerotwomix{} for different variants of fine-tuning. All models are pre-trained.}
    \label{tab:fine-tune}
    \begin{tabular}{lccccSSSS}
        \toprule
        \mrow{Model} & \multicolumn{2}{c}{fine-tune} & \mrow{Joint \\training type} & \mrow{additional \\ SI-SNR loss} & {\mrow{CER}}     & {\mrow{WER}}         & {\mrow{SDR}}        & {\mrow{SI-SNR}} \\
        \cmidrule{2-3}
         & front-end & back-end &  &  \\
        \midrule
        Conv-TasNet + RNN  & --- & --- & --- &  ---   & 13.9      & 22.9          & 13.8          & 13.4 \\
        \tbli+ fine-tune ASR       & --- & \cmark & single GPU & ---  & 6.7       & 11.7          & \ditto        & \ditto \\
        \tbli+ fine-tune TasNet    & \cmark & --- & multi GPU & --- & 7.7       & 14.2          & 10.5          & 9.5  \\
        \tblii+ SI-SNR loss & \cmark & --- & multi GPU & \cmark    & 7.7       & 14.3          & 12.5          & 12.1 \\
        \tbli+ fine-tune joint     & \cmark & \cmark & multi GPU & ---    & 6.2       & 11.7          & 9.8           & 8.4 \\
        \tblii+ SI-SNR loss & \cmark & \cmark & multi GPU & \cmark    & \thl 6.0  & 11.1          & \thl 13.8     & \thl 13.5 \\
        \tbli+ fine-tune joint TBPTT  & \cmark & \cmark & single GPU (TBPTT) & --- & 6.1       & \thl 11.0     & 11.7          & 11.5 \\
        \tblii+ SI-SNR loss & \cmark & \cmark & single GPU (TBPTT) & \cmark    & 18.0      & 23.9          & 12.4          & 12.1 \\
        \bottomrule
    \end{tabular}
\end{table*}

\newcommand{\mc}[1]{\multicolumn{1}{c}{#1}}
\newcommand\setrow[1]{\gdef\rowmac{#1}#1\ignorespaces}
\newcommand\clearrow{\global\let\rowmac\relax}
\clearrow
\begin{table*}[ht]
\vspace{-5mm}
    \centering
    \caption{Comparison of single-channel multi-speaker ASR systems. They differ heavily in their used architecture, training data and technique. $^*$: The models are evaluated on mixtures based on WSJ \cite{Seki2018_purelyendend,Chang2019_EndEndMonauralMulti} and the \glspl{WER} are not comparable to the other models.}
    \label{tab:related}
  \begin{tabular}{>{\rowmac}l>{\rowmac}c>{\rowmac}c>{\rowmac}c>{\rowmac}c>{\rowmac}c>{\rowmac}l>{\rowmac}l>{\rowmac}S>{\rowmac}S<{\clearrow}}
        \toprule
        \mrow{Model}&\mrow{structure}&\mrow{pre-\\training}& \mrow{joint \\ training} & \mrow{signal \\ reconstr.} & \mrow{no paral.\\data req.} &\multicolumn{2}{c}{data} & {\mrow{CER}} & {\mrow{WER}} \\
        
        \cmidrule{7-8}
         &  & &  & & & \mc{train} & \mc{eval} &  &  \\
        \midrule
        DPCL & &  &                                                  &\cmark & --- &\wsjzerotwomix       & & \\
        \tbli+ DNN-HMM \cite{Menne2019_AnalysisDeepClustering}      & hybrid &\cmark & --- & \cmark & --- & WSJ0                 & \wsjzerotwomix & {---} & 16.5 \\
        \tbli+ CTC/attention \cite{Settle2018_Endendmulti}& E2E & \cmark & --- & \cmark & --- & WSJ0               & \wsjzerotwomix & 23.1 & {---} \\
        \tblii+ joint fine-tuning \cite{Settle2018_Endendmulti}               & E2E & \cmark & \cmark & \cmark & --- & \wsjzerotwomix       &\wsjzerotwomix& 13.2 & {---} \\
        PIT-ASR (best) \cite{Qian2018_Singlechannelmulti,Chang2019_EndEndMonauralMulti} & hybrid & --- & \cmark & --- & \cmark & \wsjzerotwomix & \wsjzerotwomix & {---} & 28.2 \\
        E2E ASR \cite{Chang2019_EndEndMonauralMulti} & E2E & --- & \cmark            & --- & \cmark & \wsjzerotwomix       & \wsjzerotwomix & {---} & 25.4 \\
        \setrow{\itshape} E2E ASR$^*$ \cite{Seki2018_purelyendend}        & E2E  & (\cmark) &    \cmark         & --- & \cmark & \wsjtwomix           & \wsjtwomix     &   13.2   &  28.2 \\ 
        \setrow{\itshape}E2E ASR$^*$ \cite{Chang2019_EndEndMonauralMulti} & E2E & --- & \cmark             & --- & \cmark & \wsjtwomix           & \wsjtwomix & 10.9 & 18.4 \\
        \midrule
        joint TasNet (our best) & E2E & \cmark & \cmark & \cmark & --- & \begin{tabular}{@{}l@{}}WSJ \&\\ \wsjzerotwomix\end{tabular} & \wsjzerotwomix{} & 6.1 & 11.0 \\
        \bottomrule
    \end{tabular}
    \vspace{-5mm}
\end{table*}

\subsection{CTC/attention ASR model}
%
%
We use a configuration similar to \cite{Seki2018_purelyendend} without the speaker dependent layers for the speech recognizer.
This results in a model with two CNN layers followed by two BLSTMP layers with $1024$ units each for the encoder, one LSTM layer with 300 units for the decoder and a feature dimension of $80$.
The multi-task learning weight was set to $\lambda=0.2$.
We use a location-aware attention mechanism and ADADELTA \cite{Zeiler2012_ADADELTAadaptivelearning} as optimizer.
All decoding is performed with an additional word-level RNN language model.
%
%
Our ASR model achieves a WER of \SI{6.4}{\percent} on the WSJ eval92 set.

\subsection{Joint finetuning}

The results of the different fine-tuning variants are listed for comparison in \cref{tab:fine-tune}.
%
%
It is notable that combining the independently trained models (Conv-TasNet + RNN) already gives a competitive performance compared other methods (see \cref{sec:experiments:results:ind} and \cref{tab:related}).
Fine-tuning just the \gls{ASR} system (+ fine-tune ASR) can further cut the \gls{WER} almost in half from \SI{22.9}{\percent} to \SI{11.7}{\percent}.

%
%
Joint fine-tuning without a signal level loss (+ fine-tune joint), when the system is no longer constrained to transport meaningful speech between front- and back-end, can not improve much over just fine-tuning the ASR system and significantly lowers the source separation performance.
This indicates that there is enough information available for reliable speech recognition in the separated signals (i.e., retraining of the front-end is not required), but that not all information required to reconstruct speech is required for \gls{ASR}.

Using a signal-level loss (+ fine-tune joint + SI-SNR loss) can further improve the \gls{WER} to \SI{11.1}{\percent}.
In this case, the source separation performance stays comparable to the Conv-TasNet model.
A signal-level loss helps the model to better separate the speech.

%
%
The performance for just fine-tuning the front-end (+ fine-tune TasNet) cannot reach the performance of fine-tuning the back-end.
This means that it is easier to mitigate the mismatch for the \gls{ASR} back-end (i.e., learn to ignore the artifacts produced by the front-end) than it is for the front-end (i.e., learn to suppress the artifacts).

%
%
Comparing the results of the chunk-based fine-tuning (+ fine-tune joint TBPTT) as an approximation of TBPTT with the full joint fine-tuning (+ fine-tune joint), it can be seen that even though the TBPTT-based approach is just an approximation, its performance is comparable to the full joint model if no signal-level loss is used.
It even performs slightly better, possibly because TBPTT allowed to use a larger batch size.
The slightly odd degradation in performance for the case with a signal level loss (+ fine-tune joint TBPTT + SI-SNR loss) might be caused by the signal-level loss penalizing the approximation heavily, while the gradient propagated through the \gls{ASR} system is less harmful.
This case was not evaluated further.

\subsection{Comparison with related work}
\label{sec:experiments:results:ind}

%
%
This section compares the performance of the different related works presented in \cref{sec:related}.
Their major differences and performance in terms of \gls{CER} and \gls{WER} are listed in \cref{tab:related}.
While these comparisons are not fair because the presented works differ greatly in their overall model structure, training methods and data, the numbers are meant to give a rough indication of how these methods compare and how complex the design space is.
Among the related systems, the hybrid DNN-HMM system still outperforms all monolithic approaches although this system is not fine-tuned jointly.
On the other hand, the E2E ASR model can outperform the jointly optimized hybrid PIT-ASR on the same dataset.
Their \glspl{WER}, however, are far from the joint DPCL model.
Although not directly comparable, the best results in this table were produced by cascade models that allow reconstruction of the enhanced separated signals (\gls{DPCL} + ASR, joint TasNet), which suggests that having dedicated parts for source separation and speech recognition is helpful, while joint fine-tuning improves the performance.
Our time domain approach gives the best result in this comparison.

\section{Conclusions}
We propose to use a time domain source separation system like Conv-TasNet as a front-end for a single-speaker \gls{E2E} \gls{ASR} system to form a multi-speaker \gls{E2E} speech recognizer.
We show that independently training the front- and back-end already gives a competitive performance and that joint fine-tuning can drastically improve the performance.
Fine-tuning can be performed jointly with the whole model distributed over multiple GPUs, but can as well be sped up roughly by a factor of $2$ on a single GPU by approximating TPBTT for convolutional neural networks, while keeping the performance comparable.
The results suggest that retraining the \gls{ASR} part can much better compensate the mismatch between front-end and back-end than a fine-tuned front-end could.

\pagebreak
\balance
\bibliographystyle{IEEEbib}
\bibliography{library,refs}

\begin{thebibliography}{10}

\bibitem{Xiong2018_Microsoft2017ConversationalSpeech}
W.~{Xiong}, L.~{Wu}, F.~{Alleva}, J.~{Droppo}, X.~{Huang}, and A.~{Stolcke},
\newblock ``The microsoft 2017 conversational speech recognition system,''
\newblock in {\em 2018 IEEE International Conference on Acoustics, Speech and
  Signal Processing (ICASSP)}, April 2018, pp. 5934--5938.

\bibitem{ami}
J.~Carletta, S.~Ashby, S.~Bourban, M.~Flynn, M.~Guillemot, T.~Hain, J.~Kadlec,
  V.~Karaiskos, W.~Kraaij, M.~Kronenthal, G.~Lathoud, M.~Lincoln, A.~Lisowska,
  I.~McCowan, W.~Post, D.~Reidsma, and P.~Wellner,
\newblock ``The ami meeting corpus: A pre-announcement,''
\newblock in {\em Machine Learning for Multimodal Interaction}, S.~Renals and
  S.~Bengio, Eds., Berlin, Heidelberg, 2006, pp. 28--39, Springer Berlin
  Heidelberg.

\bibitem{Isik2016_Singlechannelmulti}
Y.~Isik, J.~L. Roux, Z.~Chen, S.~Watanabe, and J.~R. Hershey,
\newblock ``Single-channel multi-speaker separation using deep clustering,''
\newblock {\em arXiv preprint arXiv:1607.02173}, 2016.

\bibitem{Yu2016_PermutationInvariantTraining}
D.~Yu, M.~Kolb{\ae}k, Z.-H. Tan, and J.~Jensen,
\newblock ``Permutation invariant training of deep models for
  speaker-independent multi-talker speech separation,''
\newblock in {\em The 42nd IEEE International Conference on Acoustics, Speech
  and Signal ProcessingIEEE International Conference on Acoustics, Speech and
  Signal Processing}. IEEE, 2017, pp. 241--245.

\bibitem{Luo2017_tasnet}
Y.~Luo and N.~Mesgarani,
\newblock ``Tasnet: time-domain audio separation network for real-time,
  single-channel speech separation,''
\newblock in {\em 2018 IEEE International Conference on Acoustics, Speech and
  Signal Processing (ICASSP)}. IEEE, 2018, pp. 696--700.

\bibitem{Luo2018_ConvTasNetSurpassing}
Y.~Luo and N.~Mesgarani,
\newblock ``Conv-tasnet: Surpassing ideal time--frequency magnitude masking for
  speech separation,''
\newblock {\em IEEE/ACM Transactions on Audio, Speech, and Language
  Processing}, vol. 27, no. 8, pp. 1256--1266, 2019.

\bibitem{Menne2019_AnalysisDeepClustering}
T.~Menne, I.~Sklyar, R.~Schl{\"u}ter, and H.~Ney,
\newblock ``Analysis of deep clustering as preprocessing for automatic speech
  recognition of sparsely overlapping speech,''
\newblock {\em arXiv preprint arXiv:1905.03500}, 2019.

\bibitem{Settle2018_Endendmulti}
S.~Settle, J.~Le~Roux, T.~Hori, S.~Watanabe, and J.~R. Hershey,
\newblock ``End-to-end multi-speaker speech recognition,''
\newblock in {\em 2018 IEEE International Conference on Acoustics, Speech and
  Signal Processing (ICASSP)}. IEEE, 2018, pp. 4819--4823.

\bibitem{Yu2017_RecognizingMultitalker}
D.~Yu, X.~Chang, and Y.~Qian,
\newblock ``Recognizing multi-talker speech with permutation invariant
  training,''
\newblock {\em arXiv preprint arXiv:1704.01985}, 2017.

\bibitem{Qian2018_Singlechannelmulti}
Y.~Qian, X.~Chang, and D.~Yu,
\newblock ``Single-channel multi-talker speech recognition with permutation
  invariant training,''
\newblock {\em Speech Communication}, vol. 104, pp. 1--11, 2018.

\bibitem{Bahmaninezhad2019_comprehensivestudyspeech}
F.~Bahmaninezhad, J.~Wu, R.~Gu, S.-X. Zhang, Y.~Xu, M.~Yu, and D.~Yu,
\newblock ``A comprehensive study of speech separation: spectrogram vs waveform
  separation,''
\newblock {\em arXiv preprint arXiv:1905.07497}, 2019.

\bibitem{Kim2017_JointCTCattention}
S.~Kim, T.~Hori, and S.~Watanabe,
\newblock ``Joint {CTC}-attention based end-to-end speech recognition using
  multi-task learning,''
\newblock in {\em 2017 IEEE international conference on acoustics, speech and
  signal processing (ICASSP)}. IEEE, 2017, pp. 4835--4839.

\bibitem{Watanabe2017_HybridCTC/attentionarchitecture}
S.~Watanabe, T.~Hori, S.~Kim, J.~R. Hershey, and T.~Hayashi,
\newblock ``Hybrid {CTC}/attention architecture for end-to-end speech
  recognition,''
\newblock {\em IEEE Journal of Selected Topics in Signal Processing}, vol. 11,
  no. 8, pp. 1240--1253, 2017.

\bibitem{Chan2016_Listenattendspell}
W.~Chan, N.~Jaitly, Q.~Le, and O.~Vinyals,
\newblock ``Listen, attend and spell: A neural network for large vocabulary
  conversational speech recognition,''
\newblock in {\em 2016 IEEE International Conference on Acoustics, Speech and
  Signal Processing (ICASSP)}. IEEE, 2016, pp. 4960--4964.

\bibitem{werbos1990backpropagation}
P.~J. Werbos,
\newblock ``Backpropagation through time: what it does and how to do it,''
\newblock {\em Proceedings of the IEEE}, vol. 78, no. 10, pp. 1550--1560, 1990.

\bibitem{Seki2018_purelyendend}
H.~Seki, T.~Hori, S.~Watanabe, J.~L. Roux, and J.~R. Hershey,
\newblock ``A purely end-to-end system for multi-speaker speech recognition,''
\newblock {\em arXiv preprint arXiv:1805.05826}, 2018.

\bibitem{Chang2019_EndEndMonauralMulti}
X.~Chang, Y.~Qian, K.~Yu, and S.~Watanabe,
\newblock ``End-to-end monaural multi-speaker asr system without pretraining,''
\newblock in {\em ICASSP 2019-2019 IEEE International Conference on Acoustics,
  Speech and Signal Processing (ICASSP)}. IEEE, 2019, pp. 6256--6260.

\bibitem{Watanabe2018_ESPnet}
S.~Watanabe, T.~Hori, S.~Karita, T.~Hayashi, J.~Nishitoba, Y.~Unno, N.~E.~Y.
  Soplin, J.~Heymann, M.~Wiesner, N.~Chen, et~al.,
\newblock ``Espnet: End-to-end speech processing toolkit,''
\newblock {\em arXiv preprint arXiv:1804.00015}, 2018.

\bibitem{Heymann2017_BeamnetEndend}
J.~Heymann, L.~Drude, C.~Boeddeker, P.~Hanebrink, and R.~Haeb-Umbach,
\newblock ``Beamnet: End-to-end training of a beamformer-supported
  multi-channel asr system,''
\newblock in {\em 2017 IEEE International Conference on Acoustics, Speech and
  Signal Processing (ICASSP)}. IEEE, 2017, pp. 5325--5329.

\bibitem{Fevotte2005_BSS_EVAL}
E.~Vincent,
\newblock ``{BSSEval}. a toolbox for performance measurement in (blind) source
  separation,''
\newblock {\em l{\'\i}nea]. Disponible: http://bassdb. gforge. inria.
  fr/bss\_eval/.[{\'U}ltimo acceso: 29 Marzo 2017]}, 2005.

\bibitem{LeRoux2019_SDR}
J.~Le~Roux, S.~Wisdom, H.~Erdogan, and J.~R. Hershey,
\newblock ``{SDR}--half-baked or well done?,''
\newblock in {\em ICASSP 2019-2019 IEEE International Conference on Acoustics,
  Speech and Signal Processing (ICASSP)}. IEEE, 2019, pp. 626--630.

\bibitem{Kingma2014_Adam}
D.~P. Kingma and J.~Ba,
\newblock ``Adam: A method for stochastic optimization,''
\newblock {\em CoRR}, vol. abs/1412.6980, 2014.

\bibitem{Zeiler2012_ADADELTAadaptivelearning}
M.~D. Zeiler,
\newblock ``Adadelta: an adaptive learning rate method,''
\newblock {\em arXiv preprint arXiv:1212.5701}, 2012.

\end{thebibliography}

\end{document}